\documentclass[superscriptaddress,twocolumn,aps]{revtex4-1}

\usepackage{hyperref}
\hypersetup{
  breaklinks=true,
  colorlinks=true,
  linkcolor=blue,
  filecolor=magenta,
  urlcolor=cyan,
}

\usepackage{physics} % general physics package
\usepackage{amssymb} % math fonts and symbols
\usepackage{bm} % for making math symbols bold
\usepackage{braket} % for nice brackets (e.g. |X>)

\renewcommand{\t}{\text} % text in math mode
\newcommand{\f}[2]{\dfrac{#1}{#2}} % shorthand for fractions
\newcommand{\p}[1]{\left(#1\right)} % parenthesis
\renewcommand{\sp}[1]{\left[#1\right]} % square parenthesis
\renewcommand{\set}[1]{\left\{#1\right\}} % curly parenthesis
\newcommand{\bk}{\braket} % shorthand for braket notation

\newcommand{\x}{\text{x}}
\newcommand{\y}{\text{y}}
\newcommand{\z}{\text{z}}
\renewcommand{\d}{\text{d}}

\newcommand{\ii}{\mathrm{i}\mkern1mu} % imaginary unit
\newcommand{\ZZ}{\mathbb{Z}}
\renewcommand{\SS}{\mathbb{S}}
\newcommand{\TT}{\mathbb{T}}
\newcommand{\I}{\mathcal{I}}
\renewcommand{\P}{\mathcal{P}}

\let\var\relax
\DeclareMathOperator{\var}{var}

\usepackage{graphicx} % for figures
\graphicspath{{./figures/}} % set path for all figures
\usepackage{stackengine} % for stacking (e.g. text below figure)
\usepackage[caption=false]{subfig} % subfigures (via \subfloat[]{})
\newcommand{\sref}[1]{\protect\subref{#1}} % for referencing subfigures

\usepackage[inline]{enumitem} % in-line lists and \setlist{} (below)
\setlist[enumerate,1]{label={(\roman*)}} % default to roman numbering
\setlist{nolistsep} % more compact spacing between environments

\usepackage[dvipsnames]{xcolor}
\newcommand{\red}[1]{#1}

\usepackage{multirow}
\usepackage{hhline}

\usepackage[section]{placeins}

\begin{document}

\newcommand{\JILA}{JILA, National Institute of Standards and Technology and
  University of Colorado, 440 UCB, Boulder, Colorado 80309, USA}
\newcommand{\CTQM}{Center for Theory of Quantum Matter, University of Colorado, Boulder, CO, 80309, USA}
\newcommand{\SIT}{Department of Physics and Center for Quantum Science and Engineering, Stevens Institute of Technology, 1 Castle Point Terrace, Hoboken, NJ 07030, USA}
\newcommand{\contrib}{\thanks{Authors M.A.P.~and C.Q.~contributed equally to this work.}}

\newcommand{\thetitle}{Spin squeezing with short-range spin-exchange interactions}

\title{\thetitle}
\author{Michael A.~Perlin}
\email{mika.perlin@gmail.com}
\affiliation{\JILA}
\affiliation{\CTQM}
\author{Chunlei Qu} \contrib
\affiliation{\SIT}
\author{Ana Maria Rey}
\affiliation{\JILA}
\affiliation{\CTQM}
\date{\today}

\keywords{quantum metrology; dynamical phase transitions; phase space methods}

\begin{abstract}
We investigate many-body spin squeezing dynamics in an XXZ model with interactions that fall off with distance $r$ as $1/r^\alpha$ in $D=2$ and $3$ spatial dimensions.
In stark contrast to the Ising model, we find a broad parameter regime where spin squeezing comparable to the infinite-range $\alpha=0$ limit is achievable even when interactions are short-ranged, $\alpha>D$.
A region of ``collective'' behavior in which optimal squeezing grows with system size extends all the way to the $\alpha\to\infty$ limit of nearest-neighbor interactions.
Our predictions, made using the discrete truncated Wigner approximation (DTWA), are testable in a variety of experimental cold atomic, molecular, and optical platforms.
\end{abstract}

\maketitle

%%%%%%%%%%%%%%%%%%%%%%%%%%%%%%%%%%%%%%%%%%%%%%%%%%%%%%%%%%%%%%%%%%%%%%
{\it Introduction --}
Quantum technologies receive an enormous amount of attention for their potential to push beyond classical limits on physically achievable tasks.
In order to be useful, however, these technologies must demonstrate a practical advantage over their classical counterparts.
While most public attention has focused on a quantum advantage in the realm of computing, the quantum metrology community has made tremendous progress in developing strategies and platforms for surpassing classical limits on measurement precision \cite{giovannetti2006quantum, giovannetti2011advances, toth2014quantum, szczykulska2016multiparameter, pezze2018quantum}.
A key element in these strategies is the use of entanglement to enhance the capabilities of individual, uncorrelated quantum systems.
Spin squeezing is one of the most promising strategies for using entanglement to achieve a quantum advantage in practical sensing applications \cite{wineland1992spin, ma2011quantum}.

The paradigmatic setting for spin squeezing is the {\it one-axis twisting} (OAT) model \cite{kitagawa1993squeezed, ma2011quantum}, which generates spin-squeezed states by use of uniform, infinite-range Ising interactions that do not distinguish between the constituent spins.
These uniform interactions can be implemented directly via collisional interactions between delocalized atoms \cite{zibold2010classical, martin2013quantum, he2019engineering}, as well as indirectly through coupling to collective phonon modes \cite{britton2012engineered, bohnet2016quantum, gabbrielli2019multipartiteentanglement} or cavity photons \cite{baumann2010dicke, ritsch2013cold, norcia2018cavitymediated, kroeze2018spinor, davis2019photonmediated}.
\red{
Despite numerous proof-of-principle demonstrations, however, no spin squeezing experiment to date has achieved a practical metrological advantage, and current platforms relying on infinite-range interactions face a host of technical and fundamental difficulties that will require new breakthroughs to overcome.

The Ising model with power-law interactions that fall off with distance $r$ as $1/r^\alpha$ generates squeezing that scales with system size when $\alpha<D$ in $D$ spatial dimensions \cite{foss-feig2016entanglement}, which is highly desirable for metrological applications.
Conversely, the power-law Ising model generates only a constant amount of squeezing that is independent of system size when $\alpha>D$.
In practice, only a limited number of platforms can achieve long-range spin interactions ($\alpha<D$), making it highly desirable to shed light on the possibilities for scalable spin squeezing with short-range interactions ($\alpha>D$), which encompasses e.g.~super-exchange, dipolar, Van der Waals, and far-detuned phonon-mediated interactions.
}

Motivated by the intuition (echoed in Refs.~\cite{rey2008manybody, cappellaro2009quantum, kwasigroch2014boseeinstein, kwasigroch2017synchronization, he2019engineering, davis2020protecting}) that adding spin-exchange interactions to the Ising model should energetically protect collective behavior reminiscent of the OAT model, in this work we investigate the spin squeezing properties of the power-law XXZ model, whose ground-state physics was studied in Ref.~\cite{frerot2017entanglement}.
Remarkably, we find a broad range of parameters for which the power-law XXZ model nearly saturates the amount of squeezing generated in the infinite-range ($\alpha=0$) limit.
Even when interactions are short-ranged ($\alpha>D$), we observe a large region of collective squeezing behavior in which the amount of achievable spin squeezing grows with system size.
This region extends through to the $\alpha\to\infty$ limit of nearest-neighbor interactions.
Our work opens up \red{a new} prospect of spin squeezing in variety of cold atomic, molecular, and optical (AMO) systems, including ultracold neutral atoms \cite{cazalilla2014ultracold, gross2017quantum}, Rydberg atoms \cite{adams2019rydberg, browaeys2020manybody}, electric and magnetic dipolar quantum gasses \cite{bohn2017cold, moses2017new, lepoutre2019outofequilibrium, patscheider2020controlling}, and trapped ions \cite{britton2012engineered, bruzewicz2019trappedion}.

\begin{figure}
\centering
\stackunder{\includegraphics{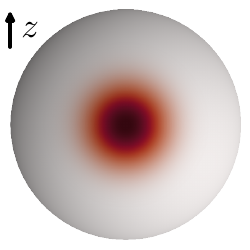}}{$t=0$}
\stackunder{\includegraphics{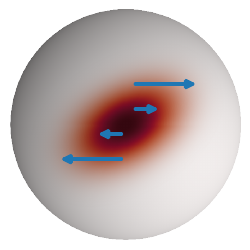}}{$t=\frac13 t_{\t{opt}}^{\t{OAT}}$}
\stackunder{\includegraphics{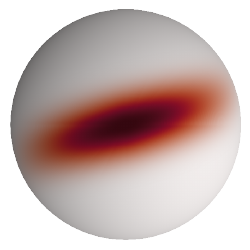}}{$t=t_{\t{opt}}^{\t{OAT}}$}
\caption{
Representations of the state $\ket{\psi(t)}$ of $N=40$ spins initially polarized along the equator, and evolved under the OAT Hamiltonian for a time $t$ up to the optimal OAT squeezing time $\chi t_{\t{opt}}^{\t{OAT}}\sim1/N^{2/3}$.
Darker colors at a point $\hat{\bm n}$ on the sphere correspond to a larger overlap $Q_{\psi(t)}\p{\hat{\bm n}}\equiv\abs{\bk{\hat{\bm n}|\psi(t)}}^2$, where $\ket{\hat{\bm n}}$ is a state in which all spins are polarized along $\hat{\bm n}$.
}
\label{fig:squeezing}
\end{figure}

%%%%%%%%%%%%%%%%%%%%%%%%%%%%%%%%%%%%%%%%%%%%%%%%%%%%%%%%%%%%%%%%%%%%%%
{\it Background and theory --}
We begin with a brief review of spin squeezing and the OAT model, described by the Ising Hamiltonian
\begin{align}
  H_{\t{OAT}} = \chi \sum_{i,j=1}^N s_{\z,i} s_{\z,j} = \chi S_\z^2,
  \label{eq:OAT}
\end{align}
where $\chi$ is the OAT squeezing strength; the spin-$z$ operator $s_{\z,i}\equiv\sigma_{\z,i}/2$ is defined in terms of the Pauli-$z$ operator $\sigma_{\z,i}$ on spin $i$; and $S_\z\equiv\sum_{i=1}^N s_{\z,i}$ is a collective spin-$z$ operator.
Eigenstates of $H_{\t{OAT}}$ can be classified by a (nonnegative) total spin $S\in\set{N/2,N/2-1,\cdots}$, and a projection $m_\z\in\set{S,S-1,\cdots,-S}$ of spin onto the $z$ axis.
The manifold of all states with maximal total spin $S=N/2$ (e.g.~spin-polarized states) is known as the {\it Dicke manifold} \cite{ma2011quantum}.
Equivalently, the Dicke manifold consists of all {\it permutationally symmetric} states that do not distinguish between underlying spins.
States in the Dicke manifold can be represented by distributions on a sphere, whose variances along different axes must satisfy an appropriate set of quantum (Heisenberg) uncertainty relations (see Figure \ref{fig:squeezing}).
In the case of a single (two-level) spin, this distribution has a fixed, Gaussian-like shape that is uniquely characterized by its orientation.
Identifying the peak of this distribution recovers the representation of a qubit state by a point on the Bloch sphere.
For $N>1$ spins, meanwhile, this distribution can acquire additional structure with metrological utility.

Given an initial state of $N$ spins polarized along the equator, represented by a Gaussian-like distribution on a sphere, the net effect of the OAT Hamiltonian is to shear this distribution, resulting in a {\it squeezed} state with a reduced variance $\p{\Delta\phi}^2$ along some axis.
This reduced variance allows for an enhanced measurement sensitivity to rotations of the collective spin state along the squeezed axis, at the expense of a reduced sensitivity to rotations along an orthogonal axis.
Spin squeezing can be quantified by the maximal gain in angular resolution $\Delta\phi$ over that achieved by a spin-polarized state \cite{ma2011quantum},
\begin{align}
  \xi^2 \equiv \f{\p{\Delta\phi_{\t{min}}}^2}{\p{\Delta\phi_{\t{polarized}}}^2}
  = \min_\phi \var(S_\phi^\perp) \times \f{N}{\abs*{\bk{\bm S}}^2},
  \label{eq:squeezing}
\end{align}
where $\bm S\equiv(S_\x,S_\y,S_\z)$ is a vector of collective spin operators; the operator $S_\phi^\perp\equiv\bm S\cdot\hat{\bm n}_\phi^\perp$ is the projection of $\bm S$ onto an axis $\hat{\bm n}_\phi^\perp$ parameterized by an angle $\phi$ in the plane orthogonal to the mean spin vector $\bk{\bm S}$; and $\var(\mathcal O) \equiv \bk{\mathcal O^2}-\bk{\mathcal O}^2$ denotes the variance of $\mathcal O$.
A spin squeezing parameter $\xi^2<1$ implies the presence of many-body entanglement \cite{sorensen2001entanglement} that enables a sensitivity to rotations beyond that set by classical limits on measurement precision \cite{giovannetti2006quantum}.
The OAT model can prepare squeezed states with $\xi^2\sim1/N^{2/3}$, whereas the fundamental (Heisenberg) limit imposed by quantum mechanics is $\xi^2\sim1/N$ \cite{giovannetti2006quantum}.

To accommodate for the fact that physical interactions are typically local, the OAT Hamiltonian in Eq.~\eqref{eq:OAT} can be modified by the introduction of coefficients $1/\abs{\bm r_i-\bm r_j}^\alpha$ in the coupling between spins $i,j$ at positions $\bm r_i,\bm r_j$, resulting in the power-law Ising model.
The introduction of non-uniform couplings means that the power-law Ising model breaks permutational symmetry, coupling the Dicke manifold of permutationally symmetric states with total spin $S=N/2$ to asymmetric states with $S<N/2$, and thereby invalidating the representation of squeezing dynamics shown in Figure \ref{fig:squeezing}.
The leakage of population outside the manifold of permutationally symmetric states can be energetically suppressed by the additional introduction of spin-aligning $\bm s_i\cdot\bm s_j$ interactions, where $\bm s_i\equiv(s_{\x,i},s_{\y,i},s_{\z,i})$ is the spin vector for spin $i$.
In total, we thus arrive at an XXZ model described by the Hamiltonian
\begin{align}
  H_{\t{XXZ}}
  = \sum_{i\ne j} \f{J_\perp \bm s_i \cdot \bm s_j + (J_\z-J_\perp) s_{\z,i} s_{\z,j}}
  {\abs{\bm r_i-\bm r_j}^\alpha}.
  \label{eq:XXZ}
\end{align}
When interactions are uniform, $\alpha=0$, the $\sum_{i\ne j}\bm s_i\cdot\bm s_j\sim \bm S^2=S\p{S+1}$ \red{term in Eq.~\eqref{eq:XXZ} is a constant of motion} within manifolds of definite total spin $S$, resulting in an OAT model with $\chi=J_\z-J_\perp$.

When $J_\z-J_\perp=0$, the XXZ model contains only the spin-aligning $\bm s_i\cdot\bm s_j$ terms, and if interactions are long-ranged, $\alpha\le D$, then the Dicke manifold is gapped away from all orthogonal states by a non-vanishing energy difference $\Delta_{\t{gap}}\gtrsim\abs{J_\perp}$ (see the Supplemental Material \red{(SM)} \cite{SM}).
As a consequence, for any finite $N$ and $\alpha\le D$ there exists a non-vanishing range of coupling strengths $J_\z\approx J_\perp$ for which a perturbative treatment of the anisotropic Ising terms in Eq.~\eqref{eq:XXZ} is valid.
In this case, the XXZ model becomes precisely the OAT model at first order in perturbation theory, with a squeezing strength $\chi_{\t{eff}} = h_\alpha \p{J_\z-J_\perp}$, where $h_\alpha$ is the average of $1/\abs{\bm r_i-\bm r_j}^\alpha$ over all $i\ne j$.
If interactions are short-ranged with $\alpha>D$, then generally $\Delta_{\t{gap}}\to0$ as $N\to\infty$, formally invalidating perturbation theory for any $J_\z$ at sufficiently large $N$.
Nonetheless, the spin-aligning terms of the XXZ model can still enable a non-perturbative emergence of ``collective'' behavior resembling perturbative, gap-protected OAT.
We numerically explore the prospect of spin squeezing with short-ranged interactions in the following section, finding that squeezing comparable to OAT may be possible with a wide range of $\alpha$ and $J_\z$, including the $\alpha\to\infty$ limit of nearest-neighbor interactions.

\begin{figure}
\centering
\includegraphics{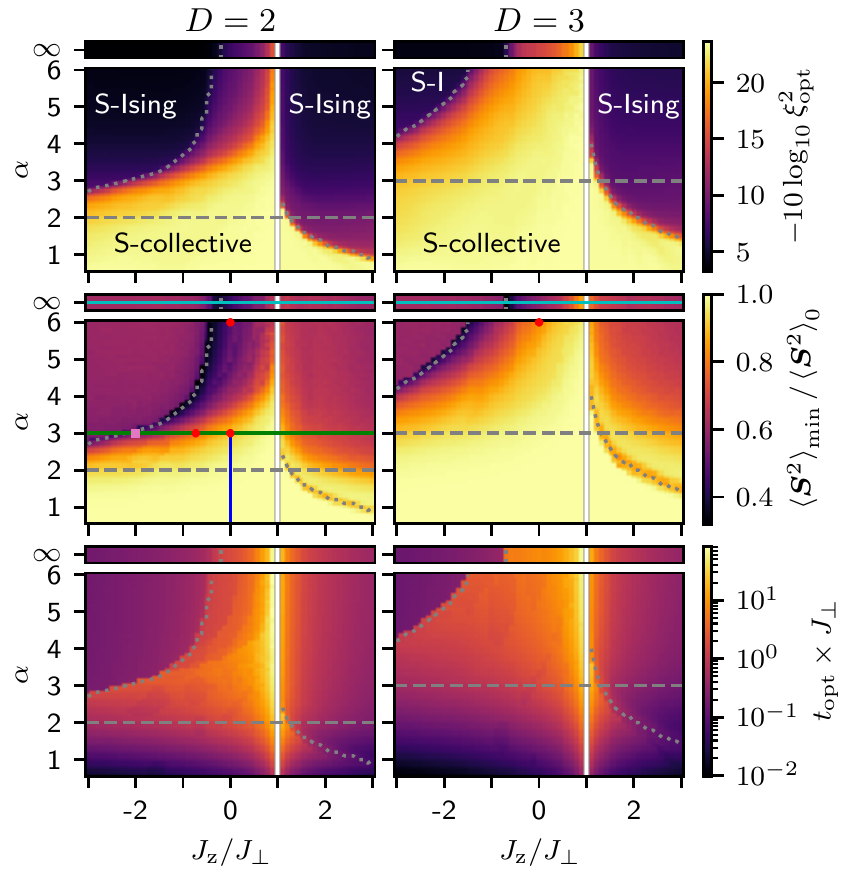}
\caption{
The optimal squeezing $\xi_{\t{opt}}^2$ (top), minimal squared \red{magnetization} $\bk{\bm S^2}_{\t{min}}$ (middle), and optimal squeezing time $t_{\t{opt}}$ (bottom) for \red{$N=4096=64^2=16^3$} spins in $D=2$ (left) and $D=3$ (right) spatial dimensions.
Spins are initially polarized along the equator and evolved under the XXZ Hamiltonian in Eq.~\eqref{eq:XXZ}.
Squeezing $\xi_{\t{opt}}^2$ is shown in decibels, and $\bk{\bm S^2}_{\t{min}}$ is normalized to its initial value $\bk{\bm S^2}_0=\frac{N}{2}\p{\frac{N}{2}+1}$.
Dashed grey lines mark $\alpha=D$, and dotted grey lines track local minima of $\bk{\bm S^2}_{\t{min}}$, marking the boundary between regions of collective and Ising-limited squeezing dynamics\red{, respectively denoted ``S-collective'' and ``S-Ising''}.
Other markers in the middle panels indicate vales of $J_\z/J_\perp,\alpha,D$ that are currently accessible with neutral atoms \cite{duan2003controlling, chen2011controlling} (cyan line), Rydberg atoms \cite{adams2019rydberg, browaeys2020manybody, signoles2020glassy} (red dots), polar molecules \cite{gorshkov2011tunable, bohn2017cold, moses2017new} (green line), magnetic atoms \cite{lepoutre2019outofequilibrium, patscheider2020controlling} (pink square), and trapped ions \cite{britton2012engineered} (blue line).
DTWA results are averaged over 500 trajectories.
}
\label{fig:main_results}
\end{figure}

%%%%%%%%%%%%%%%%%%%%%%%%%%%%%%%%%%%%%%%%%%%%%%%%%%%%%%%%%%%%%%%%%%%%%%
{\it Results --}
Whereas the quantum Ising model is exactly solvable \cite{foss-feig2013nonequilibrium, worm2013relaxation}, the XXZ model in Eq.~\eqref{eq:XXZ} is not.
We therefore investigate the spin squeezing properties of the XXZ model using the discrete truncated Wigner approximation (DTWA) \cite{schachenmayer2015manybody} for \red{$N=4096=64^2=16^3$} spins, focusing on the case of two ($D=2$) and three ($D=3$) spatial dimensions (see the \red{SM} \cite{SM} for $D=1$, where our main results are less striking but still hold).
DTWA has been shown to accurately capture the behavior of collective spin observables in a variety of settings \cite{schachenmayer2015manybody, schachenmayer2015dynamics}, and we provide additional benchmarking of DTWA for the XXZ model on \red{lattices of up to $7\times 7$ spins} in the \red{SM \cite{SM}, although it will ultimately be up to experiments to verify our findings}.
Our main results are summarized in Figure \ref{fig:main_results}, in which we explore the squeezing behavior of XXZ model in Eq.~\eqref{eq:XXZ} around the isotropic (Heisenberg) point at $J_\z=J_\perp$ by varying both $J_\z/J_\perp$ and the power-law exponent $\alpha$.
Specifically, we examine (i) the optimal squeezing parameter $\xi_{\t{opt}}^2\equiv\min_t\xi^2\p{t}=\xi^2\p{t_{\t{opt}}}$, (ii) the minimal squared \red{magnetization} throughout squeezing dynamics, $\bk{\bm S^2}_{\t{min}}\equiv\min_{t\le t_{\t{opt}}}\bk{\bm S^2}\p{t}$, and (iii) the optimal squeezing time $t_{\t{opt}}$.

First and foremost, Figure \ref{fig:main_results} confirms the theoretical argument that OAT-limited squeezing should be achievable with any power-law exponent $\alpha\le D$ for some non-vanishing range of Ising couplings, $J_\z\approx J_\perp$.
\red{Moreover, when $\alpha\le D$ we observe that} this capability persists well beyond the perturbative window with $\abs{J_\z-J_\perp}\ll\abs{J_\perp}$, covering all \red{$J_\z/J_\perp<1$} shown in Figure \ref{fig:main_results} and an increasing range of \red{$J_\z/J_\perp>1$ as $\alpha\to0$}.
Even more strikingly than the behavior at $\alpha\le D$, Figure \ref{fig:main_results} shows that squeezing well beyond the Ising limit can still achievable for a wide range of Ising couplings \red{$J_\z/J_\perp<1$} when interactions are short-ranged, $\alpha>D$.
\red{
In a nearest-neighbor XXZ model ($\alpha\to\infty$), the region $\abs{J_\z}<\abs{J_\perp}$ corresponds to the equilibrium XY phase, whereas $J_\z/J_\perp<-1$ and $J_\z/J_\perp>+1$ correspond to the equilibrium Ising ferromagnet and anti-ferromagnet phases (depending on the sign of $J_\perp$) \cite{schollwock2004quantum, frerot2017entanglement}.
The asymmetry about $J_\z=J_\perp$ in Figure \ref{fig:main_results} thus hints at an interesting connection between equilibrium physics \cite{frerot2017entanglement} and far-from-equilibrium dynamical behavior of the XXZ model (discussed further in the next section)
\footnote{In fact, when $J_\perp<0$ the S-collective region at $J_\z/J_\perp<1$ is contained {\it within} the ground-state XY phase of the power-law XXZ model \cite{frerot2017entanglement}.}.
}

Though the attainable amount of squeezing generally decreases with shorter range (increasing $\alpha$) and stronger anisotropy (decreasing \red{$J_\z/J_\perp<1$}), a region of ``collective'' squeezing behavior connected to the OAT limit persists through to the $\alpha\to\infty$ limit of nearest-neighbor interactions.
This region is reminiscent of the $\frac23D\le\alpha<D$ region of the power-law Ising model ($J_\perp=0$), in which squeezing falls short of the OAT limit, but still grows with system size \cite{foss-feig2016entanglement}.

In fact, the transition \red{between collective and Ising-limited squeezing regions, which we respectively denote ``S-collective'' and ``S-Ising'' (with an ``S-'' prefix to emphasize the role of squeezing in their characterization),} is marked by a discontinuous change in both the minimal squared \red{magnetization} $\bk{\bm S^2}_{\t{min}}$ and the optimal squeezing time $t_{\t{opt}}$, signifying the presence of a dynamical phase transition.
\red{The dynamical phases in question can be characterized by the behavior of optimal squeezing $\xi_{\t{opt}}^2$, which either scales with system size or saturates to a constant value.}
We discuss and clarify these points below.

\begin{figure}
\centering
\includegraphics{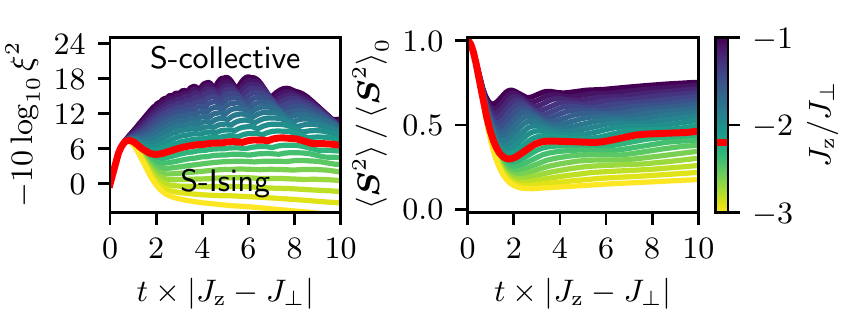}
\caption{
Squeezing $\xi^2$ and squared \red{magnetization} $\bk{\bm S^2}$ over time for the power-law XXZ model with $\alpha=3$ on a 2D lattice of $64\times64$ spins.
\red{Color indicates the value of $J_\z/J_\perp$, and red lines (at $J_\z/J_\perp=-2.2$) mark the approximate transition between S-collective and S-Ising phases, when} the ``collective'' squeezing peak at $\tau\equiv t\times \abs{J_\z-J_\perp}\sim6$ drops below the ``Ising'' peak at $\tau\sim1$.
For the parameters shown, $\bk{\bm S^2}$ reaches a minimum at $\tau\sim 2$, which means that optimal squeezing at $\tau\sim1$ is reached before maximal decay of $\bk{\bm S^2}$ in the \red{S-Ising} phase.
}
\label{fig:time_series}
\end{figure}

The discontinuity in optimal squeezing time $t_{\t{opt}}$ at the dynamical phase \red{boundary} in Figure \ref{fig:main_results} is the result of a competition between local optima in squeezing over time, shown in Figure \ref{fig:time_series}.
Large amounts of spin squeezing are generated \red{in the S-collective phase near the isotropic point at $J_\z=J_\perp$}.
The amount of squeezing generated by collective dynamics falls off away from the isotropic point, until it finally drops below an ``Ising'' squeezing peak that is generated at much short times, resulting in a discontinuous change in the time at which squeezing is optimal.
The discontinuous change in the optimal squeezing time is in turn responsible for the sudden change in the minimal squared \red{magnetization} $\bk{\bm S^2}_{\t{min}}$, which has less time to decay in the Ising-limited \red{(S-Ising)} regime.

\begin{figure}
\centering
\includegraphics{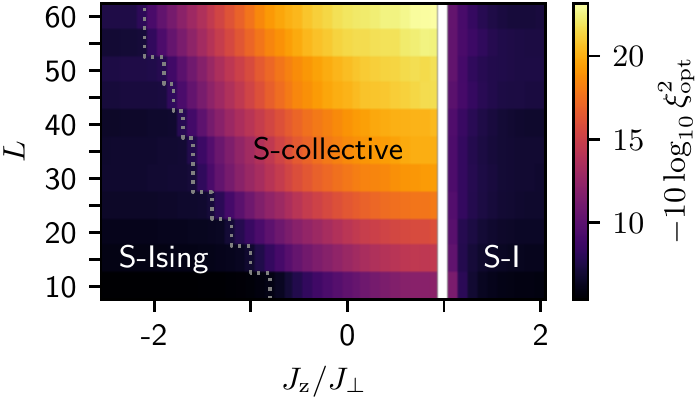}
\caption{
Optimal squeezing $\xi^2_{\t{opt}}$ as a function of system size for the power-law XXZ model with $\alpha=3$ on a 2D lattice of $N=L\times L$ spins.
Whereas the amount of squeezing generated in the \red{S-}Ising phase is insensitive to system size, \red{squeezing in the S-collective phase} grows with system size and as $J_\z/J_\perp\to1$ (from below).
Dotted grey line tracks minima of $\bk{\bm S^2}_{\t{min}}$ as a function of $J_\z/J_\perp$, as in Figure \ref{fig:main_results}, marking the approximate dynamical phase boundary.
}
\label{fig:size_scaling}
\end{figure}

It is no surprise that quantities such as $t_{\t{opt}}$ and $\bk{\bm S^2}_{\t{min}}$ that are defined via minimization exhibit discontinuous behavior, and these discontinuities do not by themselves indicate a transition between different phases of matter.
\red{We can formally distinguish the S-collective and S-Ising phases by examining the nature of squeezing that is generated in these regions.}
Specifically, the \red{S-Ising} phase generates an amount of squeezing that is insensitive to system size, whereas the \red{S-collective} phase generates an amount of squeezing that scales with system size as $\xi_{\t{opt}}^2\sim1/N^\nu$, where the exponent $\nu$ \red{generally} depends on the values of $\alpha$ and $J_\z/J_\perp$ \cite{SM}.
\red{Numerically, we find that the S-collective phase spans all $J_\z/J_\perp<1$ when $\alpha\lesssim D$, whereas the transition between S-collective and S-Ising phases occurs at a critical Ising coupling $J_\z^{\t{crit}}$ that either diverges logarithmically with system size ($J_\z^{\t{crit}}\sim-\log N$) or stays constant when $\alpha\gtrsim D$ (see Figure \ref{fig:size_scaling}, where we focus on $D=2$ and $\alpha=3$ due to its experimental relevance, and the SM \cite{SM}).}
We note that small oscillations in squeezing over time (see Figure \ref{fig:time_series}) add minor corrections to the behavior of $\xi_{\t{opt}}^2$ and $J_\z^{\t{crit}}$.
These oscillations are responsible for the discontinuous behavior of $t_{\t{opt}}$ and $\bk{\bm S^2}_{\t{min}}$ \red{seen in Figure \ref{fig:main_results} within the S-collective phase}.

\red{
{\it Discussion --}
The mechanism behind the collective dynamics featured by the XXZ model far from the isotropic point at $J_\z=J_\perp$ is not obvious, and lies in a parameter regime beyond the reach of exact treatment with current theoretical capabilities.
While an in-depth understanding of collective dynamics will most likely require experimental investigations in the spirit of quantum simulation, we discuss possible phenomenological explanations below.

Collective squeezing behavior when $\alpha<D$ is the least surprising, as the XXZ model essentially interpolates between perturbative, gap-protected OAT (near $J_\z=J_\perp$) and the long-range power-law Ising model (at $J_\z\to\pm\infty$), both of which generate collective spin squeezing.
When $\alpha>D$, as long as $D>2$ or $\alpha<2D$ (i.e.~all $\alpha>3$ when $D=3$, and $2<\alpha<4$ when $D=2$) a generalized version of the Mermin-Wagner theorem \cite{bruno2001absence} allows for the existence of long-range order in the thermodynamic limit, below a critical temperature \cite{kwasigroch2014boseeinstein, kwasigroch2017synchronization}.
Our observations may therefore be indicative of thermalization to a long-range-ordered steady state in an equilibrium XY phase \footnotemark[1], with significant amounts of collective spin squeezing present in the transient dynamics.
This explanation is supported by the fact that the squared magnetization $\bk{\bm S^2}$ approaches a non-vanishing steady-state value in Figure \ref{fig:time_series} (see also the SM \cite{SM}).
Nevertheless, the persistence of long-range order is a necessary but insufficient condition to characterize the types of dynamical phases considered in this work.
Instead, these phases are defined operationally by whether attainable spin squeezing scales with system size, and are thus sensitive to transient effects.

For even shorter range interactions ($\alpha\ge2D$) when $D\le2$, long-range order is forbidden in the steady state.
Even so, a spin-polarized initial state can still take an appreciable amount of time to thermalize to a disordered steady state.
Squeezing beyond the Ising limit can therefore occur as a transient phenomenon, before long-range order is disrupted \cite{SM}.
}

%%%%%%%%%%%%%%%%%%%%%%%%%%%%%%%%%%%%%%%%%%%%%%%%%%%%%%%%%%%%%%%%%%%%%%
{\it Experimental applications --}
As indicated in Figure \ref{fig:main_results}, our results are readily applicable to the generation of spin squeezed states in a variety of experimental platforms that have been shown to implement the power-law XXZ model, including neutral atoms ($\alpha\to\infty$) \cite{duan2003controlling, chen2011controlling}, Rydberg atoms ($\alpha=3,6$) \cite{adams2019rydberg, browaeys2020manybody, signoles2020glassy}, polar molecules ($\alpha=3$) \cite{gorshkov2011tunable, bohn2017cold, moses2017new}, and magnetic atoms ($\alpha=3$) \cite{lepoutre2019outofequilibrium, patscheider2020controlling}.
Note that one may additionally have to consider the effects of a sub-unit filling fraction on the realization of a spin model.
In principle, sub-unit filling introduces effective disorder into the XXZ spin couplings \cite{hazzard2013farfromequilibrium, kwasigroch2017synchronization}.
Nonetheless, the precise form of these interactions is not essential to the existence of \red{an S-collective} phase in the XXZ model, as evidenced by the fact that this phase persists through to the $\alpha\to\infty$ limit of nearest-neighbor interactions (see the \red{SM} \cite{SM}).

Finally, we discuss the application of our results to Ising systems without 3D spin-aligning $\bm s_i\cdot\bm s_j$ interactions, as in the case of some Rydberg atom ($\alpha=3,6$) \cite{adams2019rydberg, browaeys2020manybody} and trapped ion ($0\leq \alpha<3$) \cite{britton2012engineered} experiments.
In this case, 2D spin-aligning interactions within the $y$-$z$ plane can still be engineered by the application of a strong transverse driving field $\Omega S_\x$.
If the drive strength \red{$\abs{\Omega}\gg\frac12 N h_\alpha\abs{J_\z}$}, with $h_\alpha$ the mean of $1/\abs{\bm r_i-\bm r_j}^\alpha$ over all $i\ne j$, then moving into the rotating frame of the drive and eliminating fast-oscillating terms results in an XX model described by the Hamiltonian
\begin{align}
  H_{\t{XX}} = \f{J_\z}{2}
  \sum_{i\ne j} \f{s_{\y,i} s_{\y,j} + s_{\z,i} s_{\z,j}}{\abs{\bm r_i-\bm r_j}^\alpha},
\end{align}
which is a special case of the XXZ model in Eq.~\eqref{eq:XXZ}, with $(J_\perp,J_\z)\to(J_\z/2,0)$.
Ising systems with a strong transverse field can thus access a vertical cut along \red{$J_\z/J_\perp=0$} in Figure \ref{fig:main_results}.
\red{In a similar fashion, dynamic Hamiltonian engineering protocols \cite{choi2020robust, zhou2020quantum} can transform the Ising model into an XXZ model with any $J_\z/J_\perp\ge0$, albeit at the cost of added complexity.}

%%%%%%%%%%%%%%%%%%%%%%%%%%%%%%%%%%%%%%%%%%%%%%%%%%%%%%%%%%%%%%%%%%%%%%
\begin{acknowledgments}
We thank Andrew Lucas, Thomas Bilitewski, Sean R.~Muleady, and Jeremy T.~Young for helpful technical discussions.
This work is supported  by the DARPA DRINQs grant, the ARO single investigator award W911NF-19-1-0210, NSF grant PHY-1820885, AFOSR grant FA9550-19-1-0275, NSF grant PHY-1734006 (JILA-PFC), and by NIST.
\end{acknowledgments}

\bibliography{main.bib}

\appendix
\onecolumngrid
\section{Spectral gap of the long-range XXX model}

Here we show that the isotropic ($J_\z=J_\perp$) XXZ model in Eq.~\eqref{eq:XXZ} of the main text has a spectral gap when $\alpha\le D$, which implies the existence of a finite range of Ising couplings $J_\z\approx J_\perp$ for which the XXZ model formally recovers the OAT model at first order in perturbation theory.
For definiteness, we consider an isotropic XXZ model on a cubic lattice with periodic boundary conditions in $D$ dimensions.
The translational and SU(2) symmetries of the isotropic XXZ model on such a lattice imply that its lowest-lying excitations can be written as spin waves of the form
\begin{align}
  \ket{m_\z,k} \propto \sum_{n\in\ZZ_L^D} e^{\ii k\cdot n} s_{\z,n} \ket{m_\z},
\end{align}
where $\ket{m_\z}$ is a permutationally-symmetric Dicke state with spin projection $m_\z$ onto the $z$ axis, $n=\p{n_1,n_2,\cdots,n_D}$ indexes an individual site on the lattice of $N=L^D$ spins, and $k\in\ZZ_L^D\times2\pi/L$ is a wavenumber.
The energy of the state $\ket{m_\z,k}$ with respect to the isotropic XXZ Hamiltonian is
\begin{align}
  E_k = -J_\perp \sum_{\substack{n\in\ZZ_L^D\\\abs{n}\ne0}}
  \f{1-\cos\p{k\cdot n}}{\abs{n}^\alpha},
\end{align}
where for simplicity we work in units for which the lattice spacing is 1.
The energy $E_k$ is minimized (in magnitude) by a wavenumber that underdoes one oscillation across one axis of the lattice, e.g.~$k=\p{2\pi/L,0,0,\cdots}$, which implies a spectral gap
\begin{align}
  \Delta_{\t{gap}} = \abs{J_\perp} \sum_{\substack{n\in\ZZ_L^D\\\abs{n}\ne0}}
  \f{1-\cos\p{2\pi n_1/L}}{\abs{n}^\alpha}.
\end{align}
Letting $\epsilon\equiv2/L$, we define a rescaled domain $\SS_\epsilon=\ZZ_L/\epsilon\subset\sp{-1,1}$, and substitute $x=\epsilon n$ to get
\begin{align}
  \Delta_{\t{gap}} = \abs{J_\perp} \epsilon^{\alpha-D}
  \sum_{\substack{x\in\SS_\epsilon^D\\\abs{x}\ge\epsilon}}
  \epsilon^D \, \f{1-\cos\p{\pi x_1}}{\abs{x}^\alpha},
\end{align}
which in the thermodynamic limit $\epsilon\to0$ is well approximated by an integral that avoids an infinitesimal region at the origin,
\begin{align}
  \Delta_{\t{gap}} \to \abs{J_\perp} \epsilon^{\alpha-D} \I_D\p{\epsilon},
  &&
  \I_D\p{\epsilon} \equiv \int_{\TT_1^D\setminus\TT_\epsilon^D}
  \d^D x\, \f{1-\cos\p{\pi x_1}}{\abs{x}^\alpha},
\end{align}
where $\TT_a\equiv\p{-a,a}$ is a symmetric interval about 0.
The integrand of $\I_D\p{\epsilon}$ is strictly positive and well-behaved on the entirety of its domain except for the origin, where depending on the value of $\alpha$ the integrand may vanish or diverge as $\abs{x}\to0$.
Together, these facts mean that
\begin{align}
  \I_D\p{\epsilon} \stackrel{\epsilon\to0}{\sim} \epsilon^{-\gamma},
  &&
  \Delta_{\t{gap}} \stackrel{\epsilon\to0}{\sim} \epsilon^{\alpha-(D+\gamma)},
\end{align}
for some $\gamma\ge0$, which implies that $\Delta_{\t{gap}}>0$ when $\alpha\le D$.

%%%%%%%%%%%%%%%%%%%%%%%%%%%%%%%%%%%%%%%%%%%%%%%%%%
\section{Numerical results in one spatial dimension}

\begin{figure*}
\centering
\includegraphics{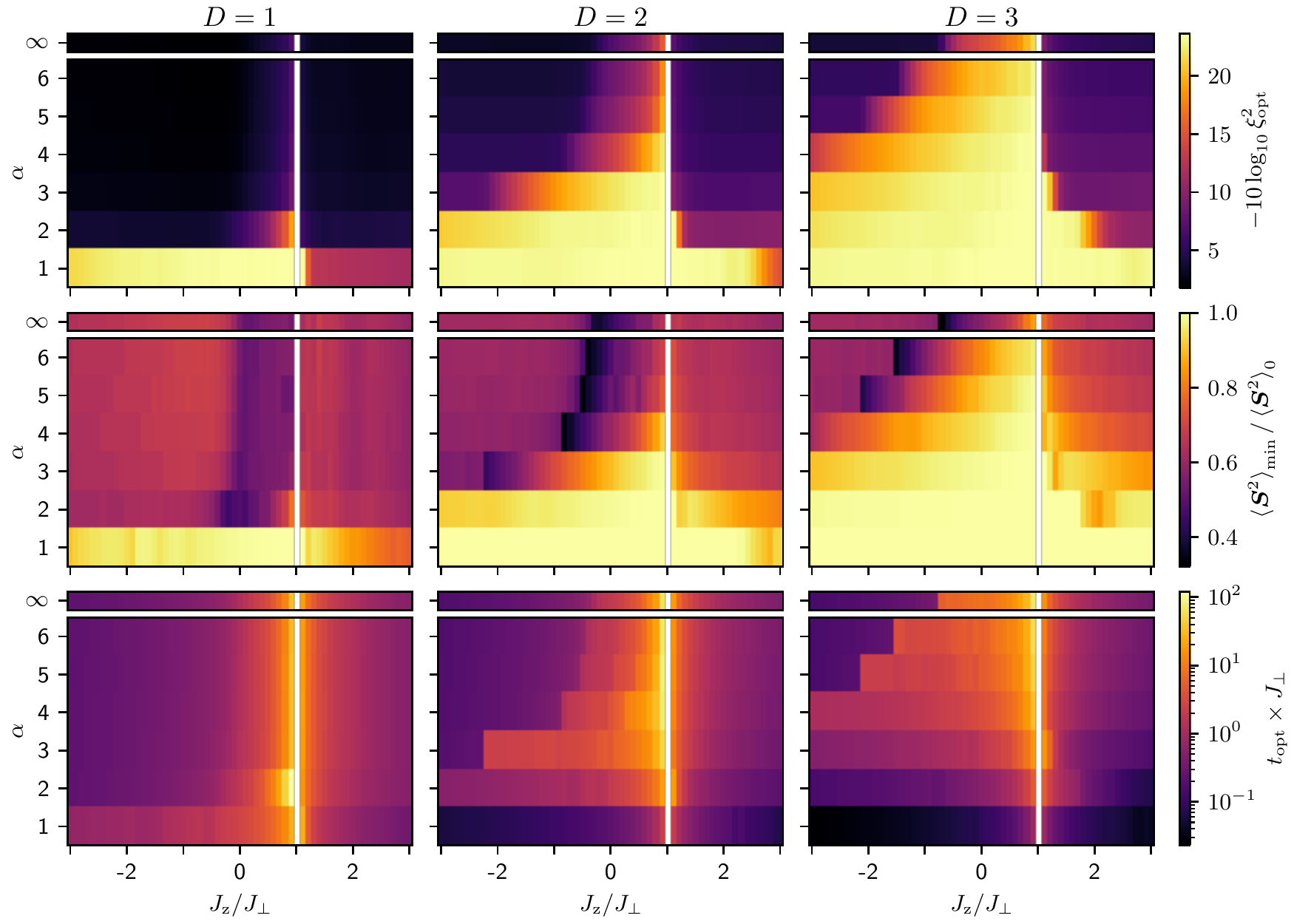}
\caption{
The optimal squeezing $\xi_{\t{opt}}^2$ (top), minimal squared \red{magnetization} $\bk{\bm S^2}_{\t{min}}$ (middle), and optimal squeezing time $t_{\t{opt}}$ (bottom) for \red{$N=4096=64^2=16^3$} spins in $D=1,2,3$ spatial dimensions.
Spins are initially polarized along the equator and evolved under the XXZ Hamiltonian in Eq.~\eqref{eq:XXZ} of the main text.
The results for $D=2$ and $3$ shown here are a subset of the results in Figure \ref{fig:main_results}, presented in the same format as that for $D=1$ for comparison.
}
\label{fig:dimensions}
\end{figure*}

Here we provide additional DTWA simulation results for the squeezing behavior of the power-law XXZ model in $D=1$ spatial dimension.
Figure \ref{fig:dimensions} shows results analogous to those in Figure \ref{fig:main_results} of the main text, for $D=1,2,3$ spatial dimensions and integer values of the power-law exponent $\alpha$ (as well as the $\alpha\to\infty$ limit of nearest-neighbor interactions).
The existence of \red{an S-collective} dynamical phase persists in one spatial dimension, but for a much narrower range of parameters than in the case of $D=2$ and $3$.
The achievable squeezing in the \red{S-collective} phase also scales less favorably with system size in the case of $D=1$.
Nonetheless, squeezing beyond the Ising limit is still achievable in $D=1$ with \red{e.g.~}$J_\z=0$ and $\alpha>1$, which is relevant for trapped ion experiments.

%%%%%%%%%%%%%%%%%%%%%%%%%%%%%%%%%%%%%%%%%%%%%%%%%%
\section{Benchmarking DTWA for the power-law XXZ model}

\begin{figure}
\centering
\includegraphics{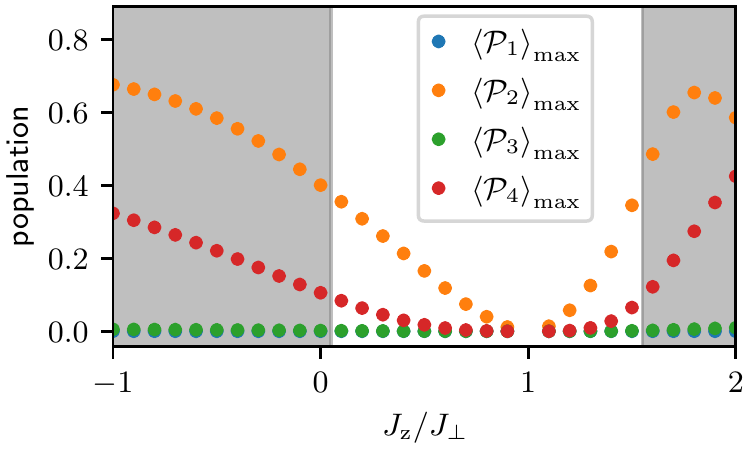}
\caption{
Maximal populations $\bk{\P_n}_{\t{max}}$ of the total spin $S=N/2-n$ manifolds $\P_n$ throughout squeezing dynamics of $7\times7$ spins, initially polarized along the equator and evolved under the XXZ Hamiltonian in Eq.~\eqref{eq:XXZ} of the main text with a power-law exponent $\alpha=3$.
Computed with TS$_4$ simulations and periodic boundary conditions.
Shaded regions indicate $\bk{\mathcal{P}_4}_{\t{max}}>0.1$, where TS$_4$ results cannot be trusted due to the likeliness of population leakage into truncated states.
All states in $\P_1$ break translational invariance, so the initial population $\bk{\P_1}_0=0$ is protected by the absence of translational symmetry-breaking terms in the Hamiltonian.
The population $\bk{\P_3}$, meanwhile, is small because $\P_3$ is only coupled to $\P_2$ and $\P_4$ by matrix elements that are $O\p{1/N}$ smaller than the couplings between $\P_0\leftrightarrow\P_2\leftrightarrow\P_4$.
}
\label{fig:shell_populations}
\end{figure}

In order to gauge the reliability of DTWA for the XXZ model in this work, we benchmark against {\it truncated shell} (TS$_4$) simulations of a $7\times7$ spin lattice whose dynamics are restricted to the subspace of $\sim N^5$ states with definite total spin $S\ge N/2-4$.
These simulations are motivated by the idea that spin-aligning $\bm s_i\cdot\bm s_j$ interactions energetically suppress the decay of total spin $S$ from its initial value of $N/2$ in a spin-polarized state.
As long as the total spin decay is small, TS$_4$ simulations should faithfully capture the dynamical behavior of a system.
\red{The restriction to small total spin decay implies that TS$_4$ simulations are only reliable near the isotropic point of the XXZ model at $J_\z=J_\perp$, and the $O(N^5)$ memory footprint of TS$_4$ means that it can only be used to simulate moderately-sized systems.
Nonetheless, TS$_4$ has the advantage over DTWA of being} ``self-benchmarking,'' in the sense that its breakdown can be diagnosed by a large population of the $S=N/2-4$ manifold, which indicates further population leakage into truncated states with $S<N/2-4$ (see Figure \ref{fig:shell_populations}).

\begin{figure}
\centering
\includegraphics{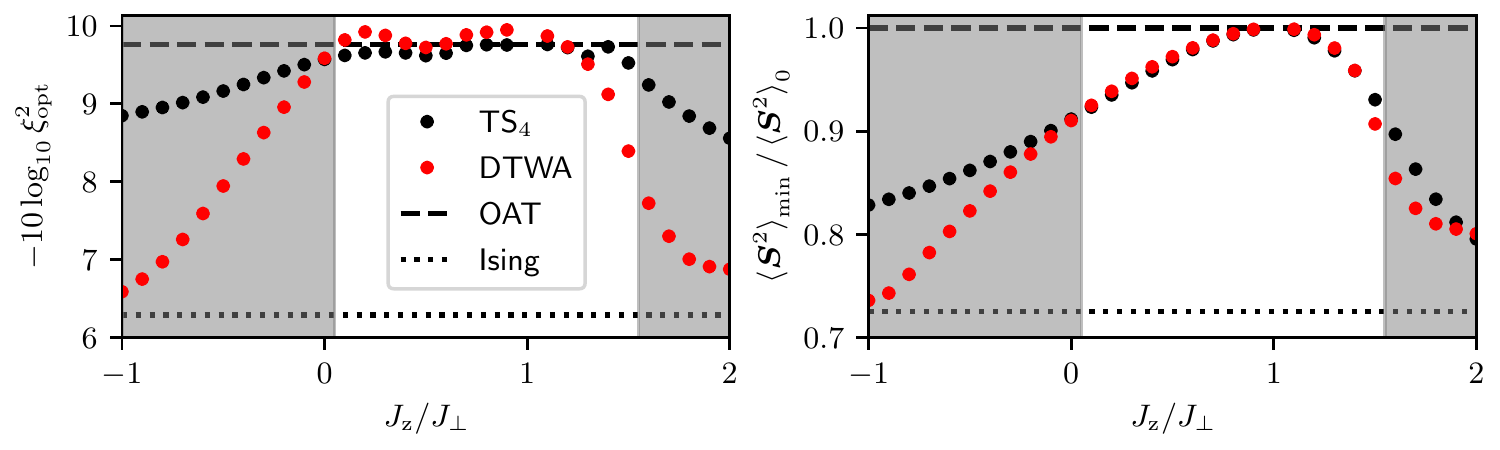}
\caption{
Optimal squeezing $\xi^2_{\t{opt}}$ (top) and minimal squared \red{magnetization} $\bk{\bm S^2}_{\t{min}}$ throughout squeezing dynamics (bottom) as computed via TS$_4$ and DTWA in the same setting as Figure \ref{fig:shell_populations}, likewise with shaded regions indicating $\bk{\mathcal{P}_4}_{\t{max}}>0.1$ in the TS$_4$ simulations.
Here squeezing $\xi_{\t{opt}}^2$ is shown in decibels, and $\bk{\bm S^2}_{\t{min}}$ is normalized to its initial value $\bk{\bm S^2}_0=\frac{N}{2}\p{\frac{N}{2}+1}$.
Dashed and dotted lines respectively mark the exactly solvable limits of uniform (OAT, $\alpha=0$) and power-law Ising (Ising, $J_\perp=0$) interactions.
}
\label{fig:benchmarking}
\end{figure}

We benchmark DTWA simulations against TS$_4$ in Figure \ref{fig:benchmarking} by comparing two observables of interest: (i)  the optimal spin squeezing parameter $\xi^2_{\t{opt}}\equiv\min_t\xi^2(t)=\xi^2(t_{\t{opt}})$, and the minimal value of $\bk{\bm S^2}$ throughout squeezing dynamics, $\bk{\bm S^2}_{\t{min}}\equiv\min_{t\le t_{\t{opt}}}\bk{\bm S^2}(t)$.
For reference, Figure \ref{fig:benchmarking} also shows the values of $\xi^2_{\t{opt}}$ and $\bk{\bm S^2}_{\t{min}}$ in the exactly solvable limits of uniform (OAT, $\alpha=0$) and power-law Ising ($J_\perp=0$) interactions.
For initially spin-polarized states, these limits have only one relevant energy scale, $J_\z-J_\perp$, so the only effect of changing $J_\z$ is to change dynamical time scales.

The results in Figure \ref{fig:benchmarking} show that DTWA agrees almost exactly with TS$_4$ in the regimes that TS$_4$ can be trusted, suggesting that DTWA is a reliable method for studying the spin squeezing behavior of the XXZ model.
Values of squeezing $-10\log_{10}\xi^2>0$ are highly sensitive to errors in collective spin observables, so \red{when comparing DTWA and TS$_4$ one should expect more pronounced (albeit minor) disagreements in spin squeezing $-10\log_{10}\xi^2$ than in squared magnetization $\bk{\bm S^2}$}.
Also, for clarity we used a simple heuristic to identify regimes of validity for TS$_4$ in Figures \ref{fig:shell_populations} and \ref{fig:benchmarking}.
This heuristic is not intended to be a precise indicator of quantitative accuracy for TS$_4$, so it is no surprise that it does not identify the precise values of $J_\z$ at which DTWA and TS$_4$ diverge.

\begin{figure}
\centering
\includegraphics{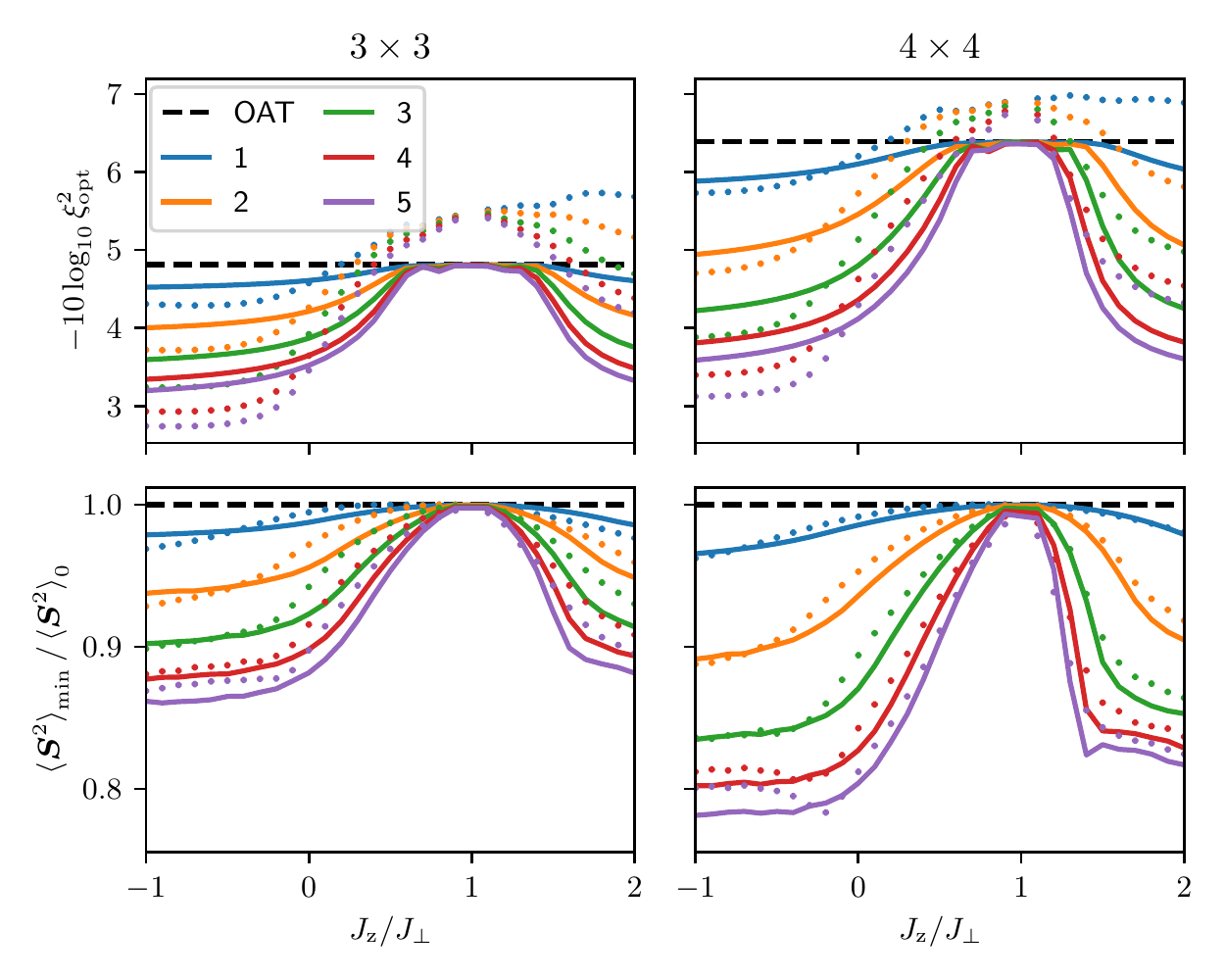}
\caption{
\red{
Optimal squeezing $\xi^2_{\t{opt}}$ (top) and minimal squared magnetization $\bk{\bm S^2}_{\t{min}}$ throughout squeezing dynamics (bottom) on 2D lattices of $3\times3$ (left) and $4\times4$ (right) spins, as computed by exact methods (solid lines) and DTWA (dots).
The color of each marker indicates the corresponding value of $\alpha$, as specified in the legend, and the dashed line marks the OAT limit of $\alpha=0$.
}
}
\label{fig:exact}
\end{figure}

\red{
Finally, Figure \ref{fig:exact} shows comparisons of DTWA with exact simulations in 2D lattices of $3\times 3$ and $4\times 4$ spins.
Though the optimal squeezing parameter $\xi_{\t{opt}}^2$ and minimal squared magnetization $\bk{\bm S^2}_{\t{min}}$ saturate to finite-size values fairly quickly away from the isotropic point at $J_\z=J_\perp$, exact simulations clearly show a collective region with OAT-limited behavior when $J_z \approx J_\perp$.
Even on small lattices, DTWA does a reasonably good job of estimating $\xi_{\t{opt}}^2$ and $\bk{\bm S^2}_{\t{min}}$.
Notably, DTWA performs better with increasing system size, as can be seen by comparing benchmarks of DTWA in $3\times3$, $4\times4$, and $7\times7$ systems, shown in Figures \ref{fig:benchmarking} and \ref{fig:exact}.
This finding is consistent with an ongoing study to benchmark DTWA against state-of-the-art simulations of matrix product states (MPS) using the time-dependent variational principle (TDVP) \cite{muleady2020dtwa}.
}

%%%%%%%%%%%%%%%%%%%%%%%%%%%%%%%%%%%%%%%%%%%%%%%%%%
\section{Scaling relations for the collective phase in $D=2$ spatial dimensions}

\begin{figure}
\centering
~\hfill
\subfloat[$\alpha=3$\label{fig:power_law_D2_a3}]
{\includegraphics{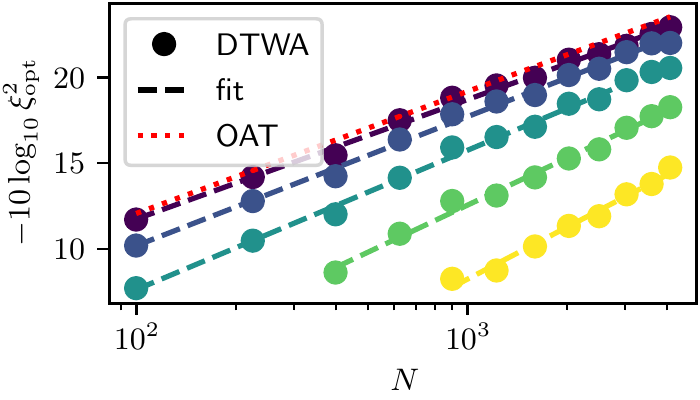}}
\hfill
\subfloat[$\alpha=4$]{\includegraphics{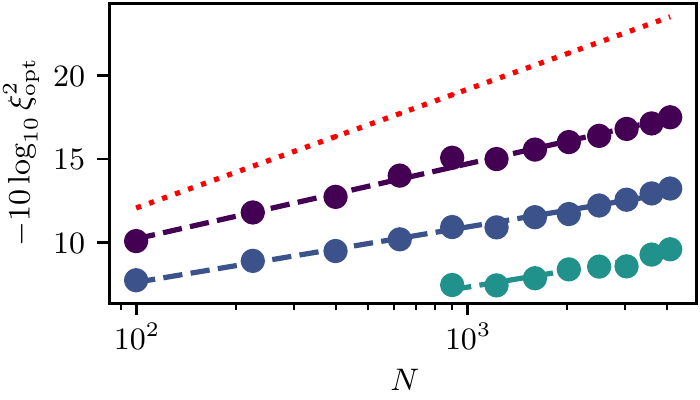}}
\hfill~\\
\hfill
\subfloat[$\alpha=5$]{\includegraphics{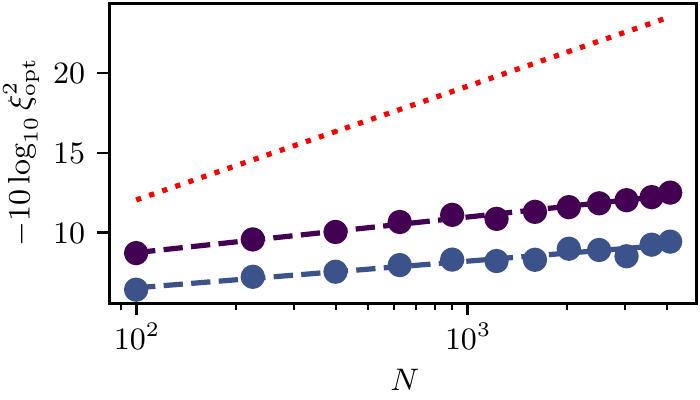}}
\hfill
\subfloat[$\alpha=6$]{\includegraphics{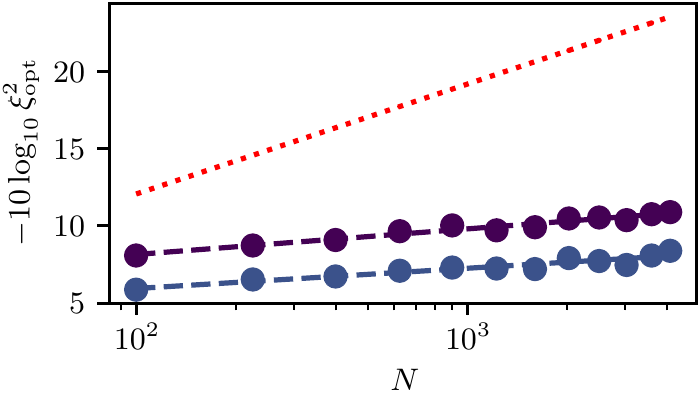}}
\hfill~\\
\subfloat[$\alpha=\infty$]{\includegraphics{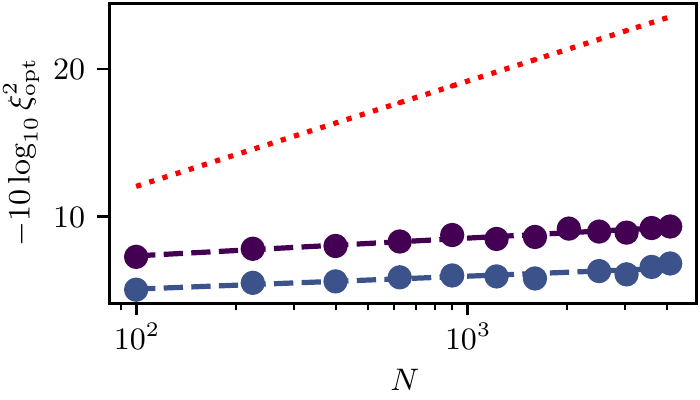}}
\caption{
Dependence of the optimal squeezing parameter $\xi^2_{\t{opt}}$ on system size $N$ within the collective dynamical phase of the power-law XXZ model in $D=2$ spatial dimensions.
Color indicates the value of $J_\z/J_\perp$, sweeping down from $+0.5$ (dark purple, top) to $-1.5$ (yellow, bottom) in increments of $-0.5$.
Circles show results computed with DTWA; dashed lines show a fit to $\xi^2_{\t{opt}}=a/N^\nu$ with free parameters $a,\nu$; and the dotted red line marks the OAT limit for reference.
The DTWA results in panel \sref{fig:power_law_D2_a3} for $\alpha=3$ are a subset of those in Figure \ref{fig:size_scaling} of the main text.
}
\label{fig:power_law}
\end{figure}

\begin{figure}
\centering
~\hfill
\subfloat[$\alpha=3$\label{fig:size_divergence_D2_a3}]
{\includegraphics{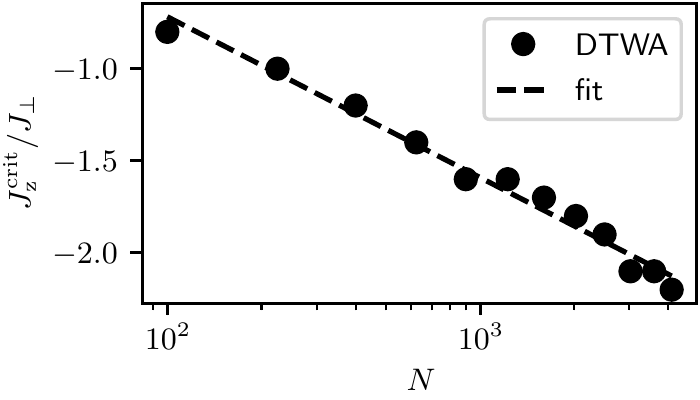}}
\hfill
\subfloat[$\alpha=4$]{\includegraphics{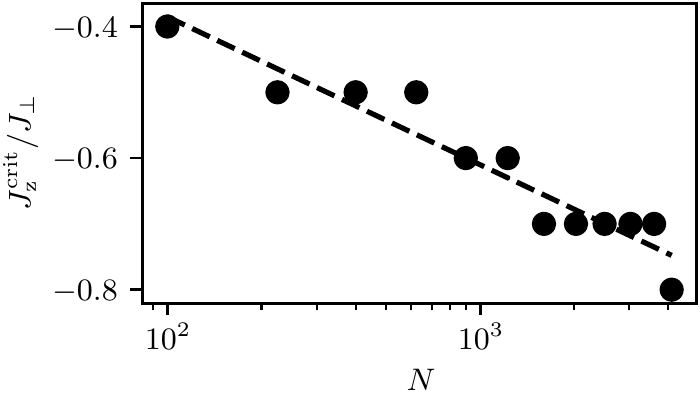}}
\hfill~\\
\hfill
\subfloat[$\alpha=5$]{\includegraphics{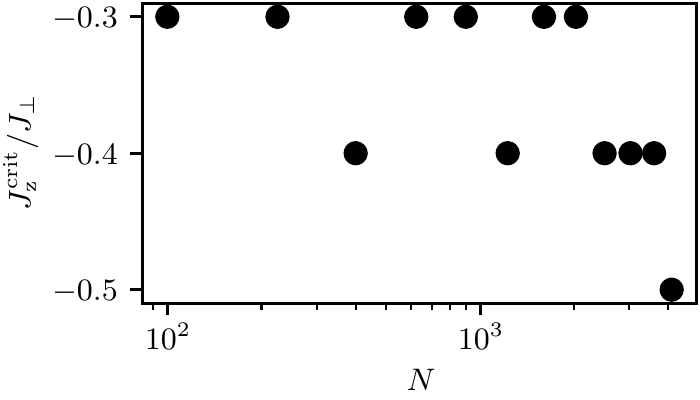}}
\hfill
\subfloat[$\alpha=6$]{\includegraphics{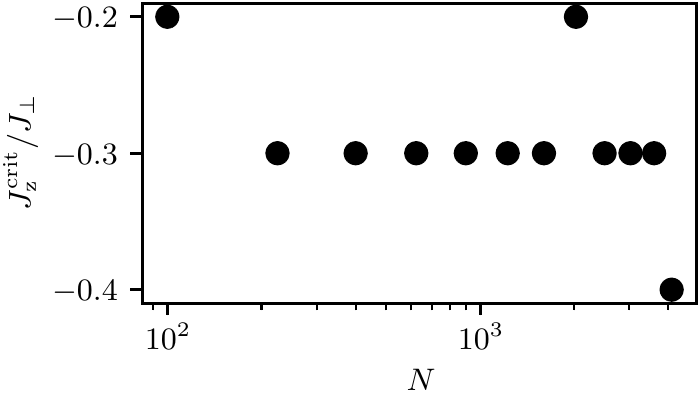}}
\hfill~\\
\subfloat[$\alpha=\infty$]{\includegraphics{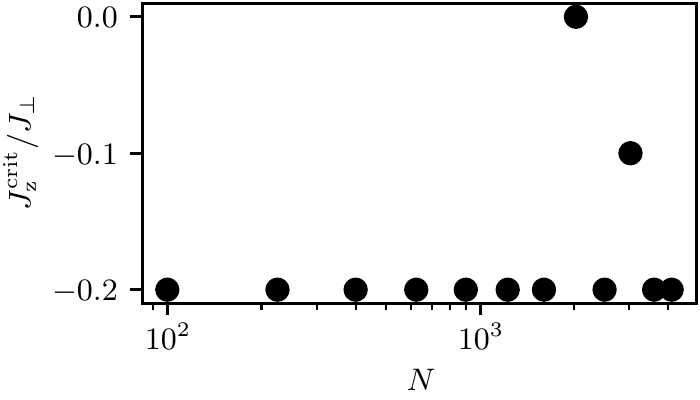}}
\caption{
Dependence of the critical Ising coupling $J_\z^{\t{crit}}$ at the collective-to-Ising dynamical phase boundary on system size $N$ for the power-law XXZ model in $D=2$ spatial dimensions.
Circles show results computed with DTWA, and dashed lines show a fit to $J_\z^{\t{crit}}/J_\perp=-\gamma\ln N+b$ with free parameters $\gamma,b$.
The DTWA results in panel \sref{fig:size_divergence_D2_a3} for $\alpha=3$ are equivalent to the dashed grey lines in Figure \ref{fig:size_scaling} of the main text.
DTWA simulations were run with values of $J_\z/J_\perp$ that are integer multiples of 0.1, placing a lower bound on the resolution for $J_\z^{\t{crit}}/J_\perp$.
}
\label{fig:size_divergence}
\end{figure}

Here we inspect the results in Figure \ref{fig:size_scaling} of the main text, as well as similar results for different exponents $\alpha$ of the power-law XXZ model, to show that
\begin{enumerate}
\item optimal squeezing scales as $\xi^2_{\t{opt}}\sim 1/N^\nu$ in the \red{S-collective} dynamical phase (Figure \ref{fig:power_law} \red{and Table \ref{tab:power_law}}), and
\item \red{the critical Ising coupling $J_\z^{\t{crit}}$ at the boundary between S-collective and S-Ising phases either diverges logarithmically with system size ($J_\z^{\t{crit}}\sim-\log N$), or remains essentially constant when $\alpha\gtrsim D$} (Figure \ref{fig:size_divergence}).
\end{enumerate}
The exponent $\nu$ governing the behavior of $\xi^2_{\t{opt}}$ will generally depend on the values of $J_\z/J_\perp$ and $\alpha$.
Similarly, the precise dependence of $J_\z^{\t{crit}}$ on $N$ will depend on the value of $\alpha$.
\red{
Note that all DTWA simulations of $N$-spin systems throughout this work average over $500\times64^2/N$ trajectories, i.e.~with 500 trajectories (samples of the initial state) for the largest system size, and $\sim1/N$ scaling to account for the fact that DTWA results converge more slowly in smaller systems.
We find that changing these trajectory numbers does not affect our overall results and conclusions.
Nonetheless, precise quantitative predictions, such as the exact value of $\nu$ as a function of system size, may be beyond our current computational capabilities, since they might require a more extensive numerical analysis to rule out finite sampling errors, or corrections from quantum correlations that are not captured by DTWA.
}

\begin{table}
\centering
\begin{tabular}{p{3ex}c||c|c|c|c|c}
  \multicolumn{2}{c}{} & \multicolumn{5}{c}{$J_\z/J_\perp$} \\[1ex]
  & & $-1.5$ & $-1.0$ & $-0.5$ & $+0.0$ & $+0.5$ \\ \hhline{~======}
  \multirow{5}{*}{$\alpha$}
  & $3$ & $1.0$ & $0.9$ & $0.8$ & $0.8$ & $0.7$ \\ \hhline{~------}
  & $4$ & -- & -- & $0.3$ & $0.3$ & $0.4$ \\ \hhline{~------}
  & $5$ & -- & -- & -- & $0.2$ & $0.2$ \\ \hhline{~------}
  & $6$ & -- & -- & -- & $0.2$ & $0.1$ \\ \hhline{~------}
  & $\infty$ & -- & -- & -- & $0.1$ & $0.1$
\end{tabular}
\caption{
\red{
Scaling exponents $\nu$ (with $\xi_{\t{opt}}^2\sim1/N^\nu$) for the values of $J_\z/J_\perp$ and $\alpha$ shown in Figure \ref{fig:power_law}, in $D=2$ spatial dimensions.
Though provided here for the sake of practical interest and transparency (these are essentially the slopes of the dashed lines Figure \ref{fig:power_law}), we note that these values are subject to correction in future work, as ruling out effects such as finite sampling errors may require a more extensive numerical analysis.
}
}
\label{tab:power_law}
\end{table}

\begin{figure}
\centering
\subfloat[$D=2$ spatial dimensions]
{\includegraphics{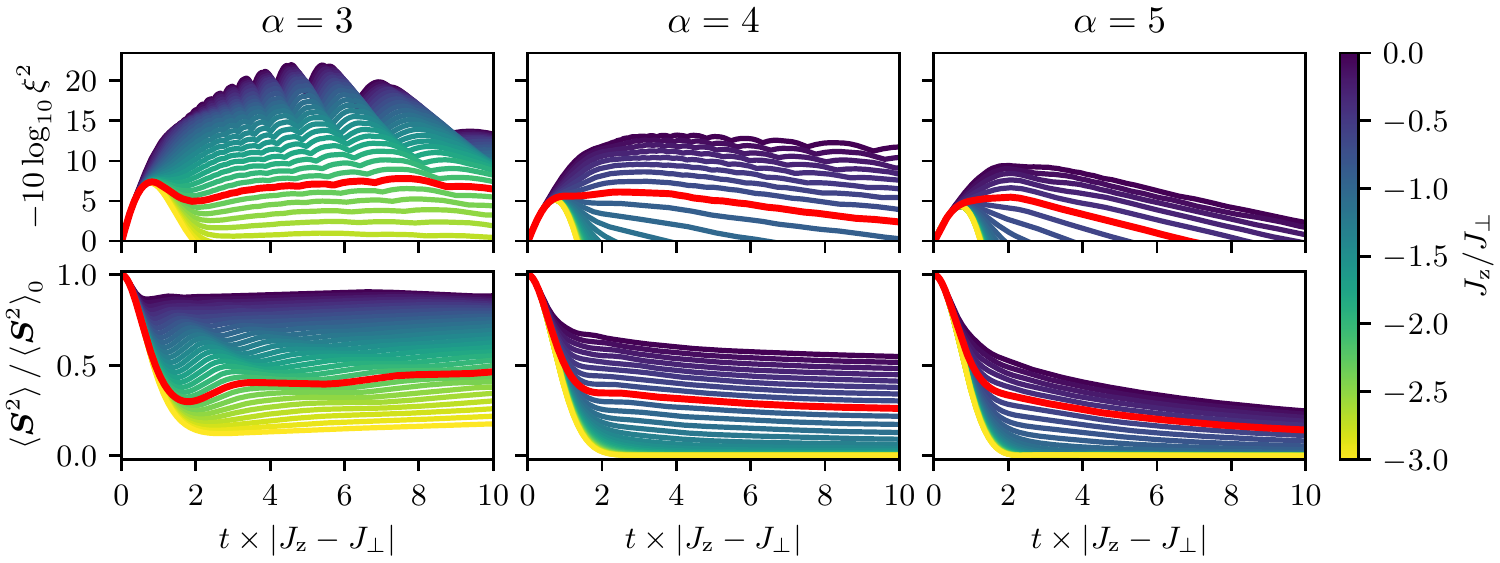}}
\\
\subfloat[$D=3$ spatial dimensions]
{\includegraphics{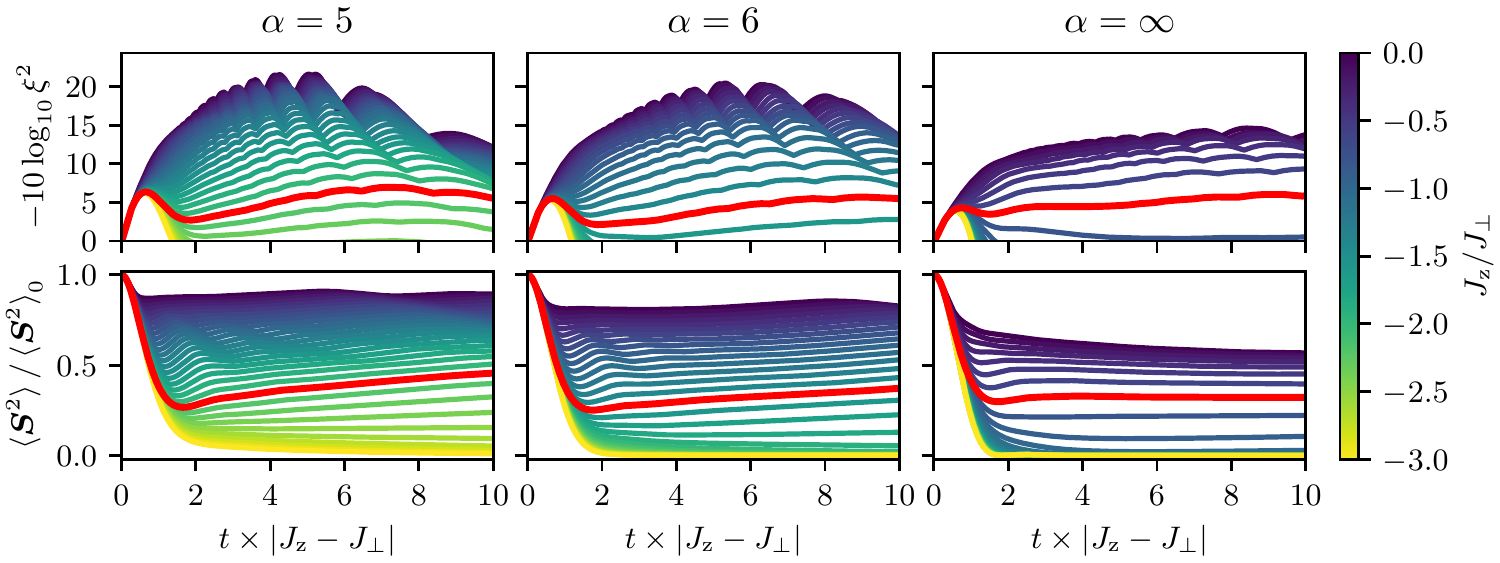}}
\caption{
\red{
Squeezing $\xi^2$ and squared magnetization $\bk{\bm S^2}$ as a function of time $t$ for $N=4096=64^2=16^3$ spins in $D=2$ and $3$ spatial dimensions.
Color indicates the value of $J_\z/J_\perp$, and the red line highlights behavior at the value of $J_\z/J_\perp$ immediately preceding the transition from the S-collective phase (above the red line) to the S-Ising phase (below the red line).
}
}
\label{fig:time_series_left}
\end{figure}

\begin{figure}
\centering
\subfloat[$D=2$ spatial dimensions]
{\includegraphics{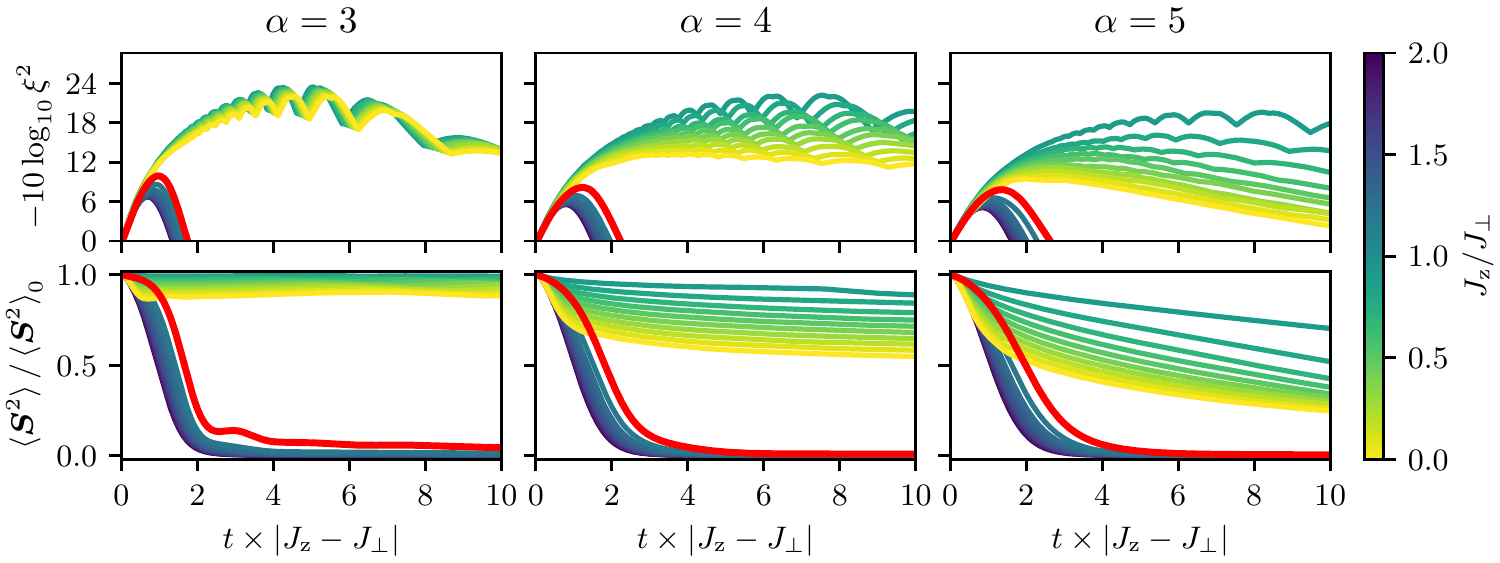}}
\\
\subfloat[$D=3$ spatial dimensions]
{\includegraphics{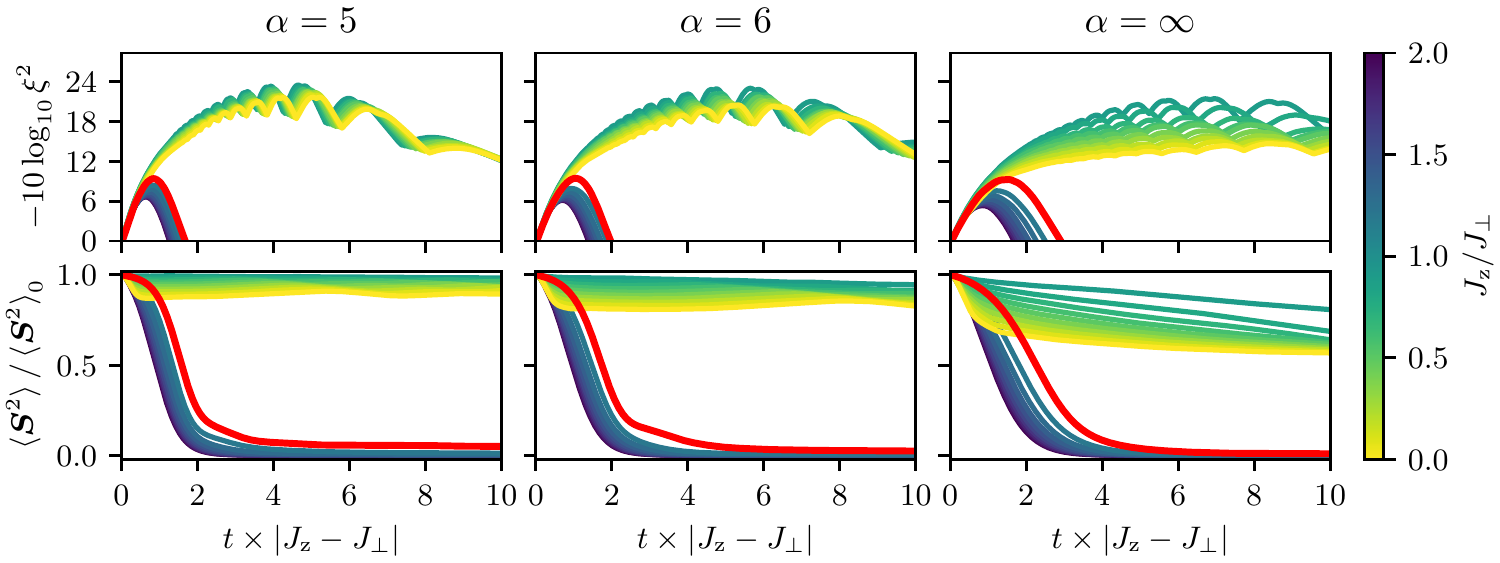}}
\caption{
\red{
Same results as in Figure \ref{fig:time_series_left}, but for values of $J_\z/J_\perp$ that cross the dynamical phase boundary at $J_\z/J_\perp=1$.
The red line highlights behavior at $J_\z/J_\perp=1.1$, immediately preceding the transition from the S-Ising phase ($J_\z>1$) to the S-collective phase ($J_\z<1$).
}
}
\label{fig:time_series_right}
\end{figure}

%%%%%%%%%%%%%%%%%%%%%%%%%%%%%%%%%%%%%%%%%%%%%%%%%%
\red{
\section{Thermalization and long-range order}

Here we provide time-series DTWA results, similar to those of Figure \ref{fig:time_series} of the main text, to show that the S-collective phase is compatible with thermalization to a long-range-ordered state of the power-law XXZ model when $D>2$ or $\alpha<2D$.
To this end, Figures \ref{fig:time_series_left} and \ref{fig:time_series_right} show both squeezing $\xi^2$ and the squared magnetization $\bk{\bm S^2}$ as a function of time for $N=4096=64^2=16^3$ spins in $D=2$ spatial dimensions with $\alpha\in\set{2D-1,2D,2D+1}=\set{3,4,5}$, as well as $D=3$ spatial dimensions with $\alpha\in\set{2D-1,2D,\infty}=\set{5,6,\infty}$.
Figure \ref{fig:time_series_left} shows simulations with values of $J_\z/J_\perp$ that sweep from $0$ (in the S-collective phase) to $-3$  (in the S-Ising phase), while Figure \ref{fig:time_series_right} shows simulations with values of $J_\z/J_\perp$ that sweep from $2$ (in the S-Ising phase) to $0$ (in the S-collective phase).
As long as $D>2$ or $\alpha<2D$, the squared magnetization $\bk{\bm S^2}$ approaches a nonzero steady-state value when $J_\z/J_\perp<1$, indicating thermalization to a steady state with long-range order.
}

%%%%%%%%%%%%%%%%%%%%%%%%%%%%%%%%%%%%%%%%%%%%%%%%%%
\section{Sub-unit filling fractions}

\begin{figure}
\centering
\includegraphics{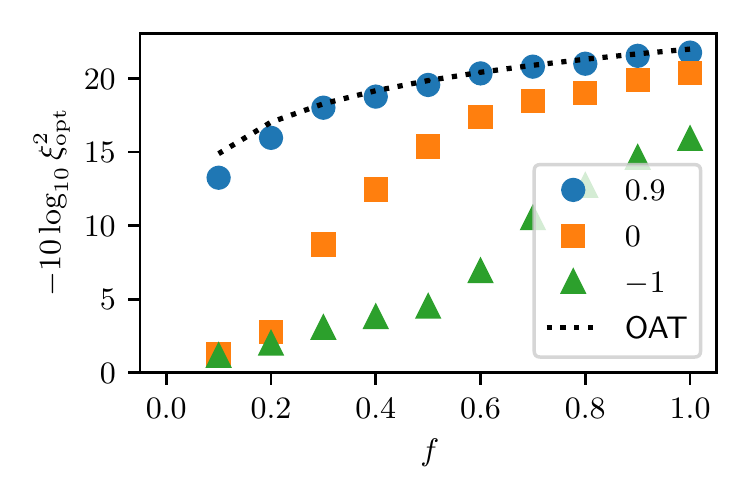}
\caption{
Dependence of the optimal squeezing parameter $\xi^2_{\t{opt}}$ on filling fraction $f$ for the XXZ model in Eq.~\eqref{eq:XXZ} of the main text with power-law exponent $\alpha=3$ in $D=2$ spatial dimensions with $50\times50$ lattice sites.
Results computed using DTWA, with a random choice of $f\times50\times50$ lattice sites to occupy.
The shape and color of each marker indicates the corresponding value of $J_\z/J_\perp$, as specified in the legend, and the dotted line marks the OAT limit for reference.
}
\label{fig:filling_fraction}
\end{figure}

Though we do not study the effect of variable filling fractions in detail, here we show that the \red{S-collective} phase is stable to filling fractions $f<1$.
To this end, in Figure \ref{fig:filling_fraction} we show the dependence of the optimal squeezing parameter $\xi^2_{\t{opt}}$ on filling fraction $f$ on a $50\times50$ lattice in $D=2$ two spatial dimensions with power-law exponent $\alpha=3$ (as in the case of polar molecules, for which unit filling is difficult to obtain experimentally).
Optimal squeezing generally decreases with filling fraction, which is in part attributable to a changing particle number.
Nonetheless, squeezing well in excess of the Ising limit is clearly achievable even for small filling fractions, $f\sim0.1$, as long as the XXZ model is tuned sufficiently close to the isotropic point at $J_\z=J_\perp$.

\red{
On a high level, decreasing the filling fraction $f$ to a value less than 1 can be seen as a two-step process: (i) rescaling all distances as $r\to r / f^{1/D}$, and (ii) adding positional disorder to spin-spin couplings, in effect transforming the XXZ Hamiltonian as
\begin{align}
  H_{\t{XXZ}} = \sum_{\substack{i\ne j\\\mu}}
  \f{J_\mu s_{\mu,i} s_{\mu,j}}{\abs{\bm r_{ij}}^\alpha}
  \to \sum_{\substack{i\ne j\\\mu}}
  \f{J_\mu s_{\mu,i} s_{\mu,j}}{\abs{\bm r_{ij}}^\alpha}
  \times f^{\alpha/D} \p{1+\epsilon_{ij}^f},
\end{align}
where the index $\mu\in\set{\x,\y,\z}$ with $J_\x=J_\y=J_\perp$, and $\epsilon_{ij}^f$ are random variables that vanish ($\epsilon_{ij}^f\to0$) as $f\to1$.
The factor $f^{\alpha/D}$ merely changes time scales, so any deviation from squeezing behavior at $f=1$ is determined by the random variables $\epsilon_{ij}^f$.
The general physics of the XXZ model at unit filling is maintained as long as these random variables are small enough to preserve the structure (connectivity) of $1/r^\alpha$ couplings.
When $f$ gets too small, however, the XXZ model is dominated by random variables, and the values of collective observables are essentially governed by the dynamics of small spin clusters with weak inter-cluster interactions.
The question remains: what filling fraction $f$ is ``too small''?

In fact, this sort of physics was studied more closely the prior work of Ref.~\cite{kwasigroch2017synchronization}, which examined the XX model ($J_\z=0$) with $1/r^3$ interactions ($\alpha=3$) and variable filling fractions that were treated as positional disorder.
By mapping this system onto one of hard-core bosons, using both mean field and numerical techniques the authors found that interactions stabilize the $1/r^3$ XX model against disorder, such that a transition from order- and disorder-dominated dynamical phases occurs at a critical filling fraction of $f_{\t{crit}}\approx0.15$.
Our results for the generic power-law XXZ model are consistent with previous results, although the role of $J_\z\ne0$ and different $\alpha$ remains an open question.
We suspect, for example, that the resilience to low filling fractions to worsen with increasing $\alpha$, and severely so when $D\le2$ and $\alpha\ge2D$, as a strengthened version of the Mermin-Wagner theorem \cite{bruno2001absence} only allows for long-range order when $D>2$ or $\alpha<2D$.
Either way, we defer a thorough analysis of this question to future work, for now merely highlighting the robustness of our main results to sub-unit filling fractions.
}

%%% Local Variables:
%%% mode: latex
%%% TeX-master: "supplement"
%%% End:

\end{document}